\documentclass[aps,prl,a4paper,reprint]{revtex4-1}

\usepackage{dsfont}
\usepackage{amsmath}
\usepackage{amssymb}
\usepackage{amsthm}
\usepackage{mathrsfs}
\usepackage{units}

\newcommand{\abs}[1]{\ensuremath{|#1|}}
\newcommand{\absbig}[1]{\ensuremath{\big|#1\big|}}

\newcommand{\normbig}[2]{\ensuremath{\Big|\!\Big|#1\Big|\!\Big|_{#2}}}

\newcommand{\tr}{\textnormal{tr}}
\newcommand{\trace}[1]{\ensuremath{\tr (#1)}}
\newcommand{\ptr}[1]{\textnormal{tr}_{\textnormal{\tiny #1}}}
\newcommand{\ptrace}[2]{\ensuremath{\ptr{#1} (#2)}}

\newcommand{\supp}[1]{\textnormal{supp}\, \{ #1 \}}

\newcommand{\ket}[1]{| #1 \rangle}

\newcommand{\bracket}[3]{\langle #1 | #2 | #3 \rangle}
\newcommand{\proj}[2]{| #1 \rangle\!\langle #2 |}

\newcommand{\kron}{\otimes}
\newcommand{\eps}{\varepsilon}

\newcommand{\cE}{\mathcal{E}}

\newcommand{\idi}[1]{\ensuremath{\mathds{1}_{\textnormal{\tiny #1}}}}
\newcommand{\idA}{\idi{A}}
\newcommand{\idX}{\idi{X}}
\newcommand{\idZ}{\idi{Z}}

\newcommand{\idx}[2]{{#1}_{\textnormal{\tiny #2}}}
\newcommand{\rhot}{\tilde{\rho}}
\newcommand{\rhoA}{\idx{\rho}{A}}
\newcommand{\rhoB}{\idx{\rho}{B}}

\newcommand{\rhoAB}{\idx{\rho}{AB}}
\newcommand{\rhoXB}{\idx{\rho}{XB}}
\newcommand{\rhotXB}{\idx{\rhot}{XB}}
\newcommand{\rhoZC}{\idx{\rho}{ZC}}
\newcommand{\rhotZC}{\idx{\rhot}{ZC}}
\newcommand{\rhotAB}{\idx{\rhot}{AB}}
\newcommand{\rhoABC}{\idx{\rho}{ABC}}
\newcommand{\rhoXXpABC}{\idx{\rho}{XX$'\!$ABC}}
\newcommand{\rhoZZpABC}{\idx{\rho}{ZZ$'\!$ABC}}
\newcommand{\rhoZZpAB}{\idx{\rho}{ZZ$'\!$AB}}
\newcommand{\rhotXXpABC}{\idx{\rhot}{XX$'\!$ABC}}
\newcommand{\rhotZZpABC}{\idx{\rhot}{ZZ$'\!$ABC}}
\newcommand{\rhoABCD}{\idx{\rho}{ABCD}}
\newcommand{\sigmaB}{\idx{\sigma}{B}}
\newcommand{\sigmaZpAB}{\idx{\sigma}{Z$'\!$AB}}
\newcommand{\tauA}{\idx{\tau}{A}}
\newcommand{\tauB}{\idx{\tau}{B}}
\newcommand{\tauAB}{\idx{\tau}{AB}}

\newcommand{\hh}[4]{\ensuremath{H_{#1}^{#2}(\textnormal{#3})_{#4}}}
\newcommand{\chh}[5]{\ensuremath{H_{#1}^{#2}(\textnormal{#3}|\textnormal{#4})_{#5}}}
\newcommand{\chmin}[3]{\chh{\textnormal{min}}{}{#1}{#2}{#3}}
\newcommand{\chmineps}[3]{\chh{\textnormal{min}}{\eps}{#1}{#2}{#3}}
\newcommand{\chmax}[3]{\chh{\textnormal{max}}{}{#1}{#2}{#3}}
\newcommand{\chmaxeps}[3]{\chh{\textnormal{max}}{\eps}{#1}{#2}{#3}}
\newcommand{\chvn}[3]{\chh{}{}{#1}{#2}{#3}}

\newcommand{\hx}[3]{\hh{#1}{}{#2}{#3}} 

\newcommand{\epsclose}{\ensuremath{\approx_\eps}}

\theoremstyle{plain}
\newtheorem{theorem}{Theorem}

\begin{document}

\title{The Uncertainty Relation for Smooth Entropies} 
\date{March 16, 2011}

\author{Marco Tomamichel}
\email{marcoto@phys.ethz.ch}
\author{Renato Renner}
\affiliation{Institute for Theoretical Physics, ETH Zurich, 8093
  Zurich, Switzerland}

\begin{abstract}
  Uncertainty relations give upper bounds on the accuracy by which the 	outcomes of two incompatible measurements can be predicted.
  While established uncertainty relations apply to cases where the
  predictions are based on purely classical data (e.g., a
  description of the system's state before measurement), an
  extended relation which remains valid in the presence of quantum
  information has been proposed recently [Berta \textit{et al.}, Nat.\
  Phys.~\textbf{6}, 659 (2010)]. Here, we generalize this uncertainty 
  relation to one formulated in terms of smooth entropies. Since these 
  entropies measure operational quantities such as extractable secret key 
  length, our uncertainty relation is of immediate practical use. To 
  illustrate this, we show that it directly implies security of a family of quantum key distribution protocols including BB84. Our proof remains valid even if the measurement devices used in the experiment deviate arbitrarily from the theoretical model.
  \end{abstract}


\maketitle


\emph{Introduction.}|\,Quantum mechanics has the peculiar property that, even if the state of
a system is fully known, certain measurements will result in a random
outcome. In other words, the information contained in the description
of a system's state is generally not sufficient to predict measurement
outcomes with certainty.
Heisenberg's uncertainty principle~\cite{heisenberg27} can be seen as a quantitative
characterization of this property.

We consider a quantum system, A, and two positive operator valued 
measurements (POVMs) acting on it, $\mathbb{X}$~with elements $\{ M_x \}$, and $\mathbb{Z}$~with elements $\{ N_z \}$.
In its entropic version, as first proposed by Deutsch and later proved 
by Maassen and Uffink~\cite{maassen88} and Krishna 
\emph{et al.}~\cite{krishna01}, the uncertainty principle reads
\begin{equation} 
	\label{eqn:class}
	\chvn{X}{S}{} + \chvn{Z}{S}{} \geq q \, .
\end{equation}
$H$ denotes the Shannon or von Neumann entropy and characterizes the uncertainty about the measurement outcomes X of $\mathbb{X}$ or Z of $\mathbb{Z}$ given any classical description, S, of the state of A before measurement~\cite{renes09}. (The most general classical description of A is a full characterization of its density matrix.)
The bound, $q$, quantifies the ``incompatibility'' of the two 
measurements and is independent of the state of A before measurement~\footnote{The norm $|\!| \cdot |\!|_{\infty}$ evaluates the largest singular value. If the measurements are projective and rank 1, namely if $M_{x} = | x \protect \rangle \! \protect \langle x |$ and $N_z = | z \protect \rangle \! \protect \langle z |$, then~\eqref{eqn:overlap} reduces to the maximum overlap, $c = \max_{x, z} | \protect \langle x | z \protect \rangle |^2$}:
\begin{align}
  q := \log_{2} \frac{1}{c} \, , \ \textnormal{where} \ \ c &:= \max_{x, z}\,
  \normbig{ \sqrt{M_x} \sqrt{N_z} }{\infty}^2 \, . \label{eqn:overlap}
\end{align}

One may now consider an agent, who, instead of holding a
classical description S of A, has access to a quantum system, B, which is fully entangled with A. It is easy to verify that this agent can predict
the outcome of any possible orthogonal measurement applied to A by performing a suitable measurement on his share of the entangled state. In
other words,\,\eqref{eqn:class} is not valid in such a generalized scenario. However, as first conjectured by Renes and Boileau~\cite{renes09}, and later proved by Berta \emph{et al.}~\cite{berta10} and Coles \emph{et al.}~\cite{coles10}, the relation
\begin{equation}
 	\label{eqn:child}
	\chvn{X}{B}{} + \chvn{Z}{C}{} \geq q \, 
\end{equation}
holds in general, for two disjoint, not necessarily classical, systems
B and C. If both systems contain only a classical description S of the
state on A, we recover~\eqref{eqn:class}~\footnote{Note that classical
  (unlike quantum) information can be ``copied'' and therefore be
  stored in two disjoint subsystems, B and C.}.

To make the above statements more precise, let $\rhoABC$ be any quantum state on three systems
A, B and C. After measuring A with respect to $\mathbb{X}$ and storing
the outcome in a classical register, X, the joint state of X and the
system B is given by~\footnote{We omit identity operators
  whenever their presence is implied by context, e.g.\ $M_x \rhoABC$ 
  should be understood as $(M_x \kron \idi{BC}) \rhoABC$.}
\begin{align*}
  \rhoXB := \sum_{x} \proj{x}{x} \kron \tauB^{x}, \ \
  \textrm{where} \ \
  \tauB^{x} = \ptr{AC} \big( M_x\, \rhoABC \big) \, .
\end{align*}
(The possible measurement outcomes of $\mathbb{X}$ are encoded in an orthonormal basis $\{ \ket{x} \}$ 
 and the probability of measuring $x$ is given by $\trace{\tauB^{x}}$.) 
Similarly, we define $\rhoZC$, where the measurement $\mathbb{Z}$
instead of $\mathbb{X}$ is applied to A and where we keep system C
instead of B. The conditional
von Neumann entropies in~\eqref{eqn:child} are then evaluated for these states, i.e.\ $\chvn{X}{B}{} = H(\rhoXB) - H(\rhoB)$.

The main contribution of this work is to generalize~\eqref{eqn:child}
to \emph{smooth entropies}~\cite{rennerwolf05,renner05}, which are
generalizations of the von Neumann entropy. Crucially, in contrast to
the latter, they characterize operational quantities beyond the
standard i.i.d.\ scenario~\footnote{Most results involving the von
  Neumann entropy are only valid for processes that produce a sequence
  of identical and independently distributed (i.i.d.) random values
  (see~\cite{renner05} for a discussion).}. For example, the smooth
min-entropy of a random variable~X conditioned on a system~B, denoted
$\chmineps{X}{B}{}$, corresponds to the number of bits contained in~X
that are $\eps$-close to uniformly distributed and independent of the
quantum system B, where $\eps \geq 0$ is the \emph{smoothing
  parameter}. Similarly the smooth max-entropy of~Z conditioned on~C,
denoted $\chmaxeps{Z}{C}{}$, corresponds to the number of bits that
are needed in order to reconstruct the value~Z using the quantum
system~C up to a failure probability~$\eps$.

The generalized uncertainty relation reads
\begin{align}
  \chmineps{X}{B}{} + \chmaxeps{Z}{C}{} \geq q \, .
  \label{eqn:mother}
\end{align}
It implies most existing uncertainty relations for two incompatible measurements~\cite{wehner09}.
In particular, it generalizes and strengthens an uncertainty relation derived via operational interpretations of the smooth entropies~\cite{renes10}. We recover~\eqref
{eqn:child} by applying the entropic asymptotic equipartition property~\cite
{tomamichel08} to~\eqref{eqn:mother}. Moreover, for $\eps = 0$ and 
disregarding B and C, we find a generalization to POVMs
of a result by Maassen and Uffink~\cite{maassen88}, bounding the
uncertainty in terms of R\'enyi entropies~\cite{renyi61} of order~$\nicefrac{1}{2}$ and $\infty$, namely $\hx{\infty}{X}{} + \hx{\nicefrac{1}{2}}{Z}{} \geq q$.

The uncertainty principle has provided intuition for various
applications, in particular in cryptography. However, previous
uncertainty relations could not be applied directly, since the von
Neumann entropy is often not the relevant measure of
uncertainty. (See~\cite{berta10} for examples and a discussion.) Our
uncertainty relation overcomes this limitation. Potential areas of
application include entanglement witnessing, the bounded storage
model~\cite{damgaard07} and quantum cryptography in general. 

As an example, we show that the relation naturally leads to a concise
and general security proof for quantum key distribution
(QKD)~\cite{bb84,ekert91}. When applied to practical prepare-and-measure protocols, it yields a strictly stronger security claim than previously known proofs. In particular, non-trivial security bounds can be obtained for realistic choices of the parameters (such as the number of exchanged signals). In addition, these bounds do not depend on the details of the measurement devices and are therefore maximally robust against imperfections in their implementation.

\emph{Smooth Entropies.}|\,For our purposes, quantum states are positive semi-definite operators
with trace smaller or equal to $1$ on a finite-dimensional Hilbert
space. Given a
state $\rhoA$ on (a Hilbert space) A, we say that $\rhoAB$ extends $\rhoA$ on B if $\ptrace{B}{\rhoAB} = \rhoA$. 
A purification is an extension of rank~$1$. We write  $\rho \epsclose
\tau$ if the purified distance between $\rho$ and $\tau$ (which is
defined as the minimum trace distance between purifications of $\rho$
and $\tau$; see \cite{tomamichel09} for details) does not exceed~$\eps$.

We now define the smooth min- and max-entropy.
Let $\eps \geq 0$ and $\rhoAB$ be a bipartite state on A and B. The
min-entropy of A given B is defined as
\begin{align*}
	\chmin{A}{B}{\rho} := \max_{\sigmaB}\ \sup \big\{
  \lambda \in \mathbb{R} :  2^{-\lambda}\, \idA \kron \sigmaB \geq
  \rhoAB \big\} \, ,
\end{align*}
where $\sigmaB$ is maximized over all states on B and $\idA$ is the
identity operator on A. Furthermore, the $\eps$-smooth min-entropy is
defined as $\chmineps{A}{B}{\rho} := \max_{\rhot}
\chmin{A}{B}{\rhot}$, where the optimization is over all states
$\rhotAB \epsclose \rhoAB$.

The smooth max-entropy is its dual~\cite{koenig08,tomamichel09} 
with regards to any purification $\rhoABC$ of $\rhoAB$ in the
sense that 
\begin{align}
  \chmaxeps{A}{B}{\rho} := - \chmineps{A}{C}{\rho} \, .
  \label{eqn:dual}  
\end{align}

We are now ready to restate our uncertainty relation.
\begin{theorem}
  \label{thm:mother}
  Let $\eps \geq 0$, let $\rhoABC$ be a tri-partite quantum state and
  let $\mathbb{X}$ and $\mathbb{Z}$ be two POVMs on A. Then,
  \begin{align*}
    \chmineps{X}{B}{\rho} + \chmaxeps{Z}{C}{\rho} \geq q\, ,
  \end{align*}
  where the entropies are evaluated using
  $\rhoXB$ and $\rhoZC$, respectively, and $\rhoXB$, $\rhoZC$ 
  and $q$ are defined as above.
\end{theorem}

\emph{Proof of the Main Result.}|\,It will be helpful to describe the two measurements in the Stinespring
dilation picture as isometries followed by a partial trace.  Let $U$
be the isometry from A to A, X and X$'$ given by $U := \sum_x \ket{x}
\kron \ket{x} \kron \sqrt{M_x}$. The isometry stores two copies of the
measurement outcome in the registers X and X$'$ and the post-measurement
state in A. Analogously, $V := \sum_z \ket{z} \kron \ket{z} \kron
\sqrt{N_z}$.  Furthermore, we introduce the states
$\rhoXXpABC := U \rhoABC U^\dagger$ and $\rhoZZpABC :=
V \rhoABC V^\dagger$, of which the post-measurement states appearing in Theorem~\ref{thm:mother}, $\rhoXB$ and $\rhoZC$, are marginals.

We now proceed to prove the theorem for the special case
where $\rhoABC$ is pure and $\eps = 0$.

  The duality relation~\eqref{eqn:dual} applied to
  $\rhoZZpABC$ gives
  \begin{align}
    \label{eqn:mo1}
    \chmax{Z}{C}{\rho} + \chmin{Z}{Z$'\!$AB}{\rho} = 0 \, .
  \end{align}
  Comparing~\eqref{eqn:mo1} with the statement of the theorem, it
  remains to show that $\chmin{Z}{Z$'\!$AB}{\rho} \leq
  \chmin{X}{B}{\rho} - q$ holds. By the definition of
  the min-entropy, we have
  \begin{align}
    &\chmin{Z}{Z$'\!$AB}{\rho} \nonumber\\
    &\quad = \max_{\sigmaZpAB} \sup \{ \lambda \in \mathbb{R}
    : 2^{-\lambda}\, \idZ \kron \sigmaZpAB \geq
    \rhoZZpAB \} \nonumber \\
    &\quad \leq \max_{\sigmaZpAB} \sup \{ \lambda \in \mathbb{R}
    : 2^{-\lambda}\, c\, \idX \kron \sigmaB \geq
    \rhoXB \} \label{eqn:mo3} \\
    &\quad = \chmin{X}{B}{\rho} - q \nonumber \, ,
	\end{align}
  where, in order to arrive at~\eqref{eqn:mo3}, we need to show that
  \begin{align}
    2^{-\lambda}\, \idZ \kron \sigmaZpAB \geq
    \rhoZZpAB \implies 2^{-\lambda}\, c\, \idX \kron \sigmaB 
\geq
    \idx{\rho}{XB} \label{eqn:mo6} \, .
  \end{align}

  For this, we apply the partial
  isometry
  $W := U V^\dagger$ followed by a partial
  trace over X$'$ and A on both sides of the inequality on the left-hand
  side. This implies
  \begin{align}
    2^{-\lambda}\, \ptr{X$'\!$A} \big( W (\idZ \kron \sigmaZpAB )
    W^\dagger \big) &\geq
    \idx{\rho}{XB} \label{eqn:mo10} \, .
  \end{align}
  Moreover, substituting the definition of $W$, we find
  	\begin{align}
	  &\ptr{X$'\!$A} \big( W (\idZ \kron \sigmaZpAB ) W^\dagger
      \big) \nonumber\\
    &\quad = \sum_{x,z} \proj{x}{x} \kron \bracket{z}{ \ptr{A} \big(
        \sqrt{N_z} M_x \sqrt{N_z} \sigmaZpAB \big)
      }{z} \label{eqn:mo8} \\
		&\quad \leq c\, \idX \kron \sigmaB \label{eqn:mo9} \, .
	\end{align}
	To get~\eqref{eqn:mo8}, we used the orthonormality of 
	$\{ \ket{x} \}_{x}$ and $\{ \ket{z} \}_{z}$ as well as the cyclicity 
	of the partial trace over A. 
  	Moreover, in the last step, we used that
  \begin{align*}
    \sqrt{N_z} M_x \sqrt{N_z} = \absbig{\sqrt{N_z} \sqrt{M_x}}^2 \leq c\, 
		\idA .
  \end{align*}
	Finally, combining~\eqref{eqn:mo9}
  with~\eqref{eqn:mo10} establishes~\eqref{eqn:mo6}, concluding the
  proof for $\eps = 0$ and pure states.

Next, we generalize this proof to $\eps$-smooth entropies.  The
purified distance used in the definition of the smooth entropies has some interesting properties~\cite{tomamichel09} 
that we use in the following:
(i) Let $\cE$ be any trace non-increasing completely positive map
  (e.g.\ a partial isometry or a partial trace). Then,
  $\rho \epsclose \tau$ implies $\cE(\rho) \epsclose
  \cE(\tau)$.
(ii) Let $\rhoAB$ be a fixed extension of $\rhoA$. Then, $\rhoA
  \epsclose \tauA$ implies that there exists an extension $\tauAB$ of
  $\tauA$ that is $\eps$-close to $\rhoAB$. Furthermore, if $\rhoAB$
  is pure and $\abs{\supp{\tauA}} \leq \dim{\textrm{B}}$, then
  $\tauAB$ can be chosen pure.

Let $\rhotZC \epsclose \rhoZC$ be a state that minimizes the
smooth max-entropy, i.e.~$\chmaxeps{Z}{C}{\rho} =
\chmax{Z}{C}{\rhot}$. Using the properties of the purified distance
discussed above, we introduce a purification
$\rhotZZpABC$, a state $\rhotXXpABC := W
\rhotZZpABC W^\dagger$ and its marginal $\rhotXB$,
which are $\eps$-close to the corresponding states $\rho$.
Applying the duality relation~\eqref{eqn:mo1} as well as the argument
in~\eqref{eqn:mo3} to $\rhot$ results in
$\chmax{Z}{C}{\rhot} + \chmin{X}{B}{\rhot} \geq q$, from which the claim follows due to the maximization over close states used in the definition of the smooth min-entropy.

Finally, to generalize the result to mixed states, we
write down the uncertainty relation for a purification $\rhoABCD$ of
$\rhoABC$, i.e.\ $\chmineps{X}{B}{} + \chmaxeps{Z}{CD}{} \geq q$. The claim is now a direct consequence of the data-processing inequality~\cite{tomamichel09}
establishing $\chmaxeps{Z}{CD}{} \leq \chmaxeps{Z}{C}{}$.

\emph{Application to Quantum Key Distribution.}|\,In the following, we
consider practically relevant \emph{prepare-and-measure} schemes such
as BB84~\cite{bb84}. In these schemes, one party, called
Alice, prepares a sequence of non-orthogonal quantum states and sends
them over a public quantum channel to a second party, Bob, who
measures these states. The correlated data gathered during this first
phase of the protocol form the \emph{raw keys}, from which Alice and
Bob can then extract a final secret key by a classical post-processing
procedure (requiring only local operations and communication over an
authenticated channel).

Amid recent hacking attacks on commercial QKD
systems~\cite{lydersen10,lo10}, it is important to point out that
information-theoretic security proofs for quantum cryptography rely on
several assumptions in addition to the validity of quantum
mechanics. 1) The two parties, Alice an Bob, have access to
genuine randomness. 2) The information that leaves each lab is
restricted to what the protocol allows. 3) The measurement devices
work according to the specifications of the protocol. These
assumptions are often not satisfied by realistic implementations.

Our novel security proof allows us to partially drop Assumption~3, which
concerns Bob's measurement device. Moreover, Assumption~2
can be weakened to allow for certain imperfections of Alice's state
preparation. The proof is based on the intuition, first formalized by
Mayers~\cite{mayers96} and captured by the uncertainty relation, that
security of QKD can be derived from the fact that Alice has a choice
between two incompatible bases for state preparation. The fact that
Bob can accurately estimate the states Alice prepared in both bases
directly implies that an eavesdropper cannot. Furthermore, this
implication holds independently of how Bob obtains his data, i.e., no
assumption about Bob's measurement device is required except that it
is memoryless. 

The proof relies on two main ingredients: (i) the uncertainty relation
(Theorem~\ref{thm:mother}) and (ii) the following result that bounds
the number of secret key bits that can be extracted from raw keys by
classical post-processing. Assume that Alice and Bob hold correlated
data, {\bf X} and {\bf X$'$}, about which an adversary
may have information E. Then, Alice and Bob can employ a classical
\emph{post-processing procedure} (usually consisting of an error
correction scheme concatenated with a procedure called privacy
amplification~\cite{bennett95,rennerkoenig05}), which generates a
shared secret key of length~\cite{renesrenner10}
\begin{align} 
  \label{eqn:pp}
  \ell \approx \chmineps{{\bf X}}{E}{} - \chmaxeps{{\bf X}}{{\bf X$'$}}{} \, .
\end{align}
(This can be seen as a single-shot version of the Devetak-Winter bound~\cite{devetak05}.) In other words, the length of the key that can be generated is essentially determined by the difference between the uncertainty that the adversary has about Alice's raw key {\bf X}, measured in terms of the smooth min-entropy, and the uncertainty that Bob has about {\bf X}, measured in terms of the smooth max-entropy.

While the following arguments are rather general, we may for concreteness consider the BB84 protocol. For the purpose of the proof we use its \emph{entanglement-based} version, which implies security of the original prepare-and-measure scheme~\cite{bennett92}. 
Here, it is assumed that Alice and Bob start with an untrusted joint quantum state, $\rhoAB$, from which they extract a secret key.
This state is supposed to be a sequence of maximally entangled qubits but may, in the presence of an adversary or noise, be arbitrarily corrupted. The protocol then proceeds as follows. First, Alice and Bob both measure each of these qubits with respect to a basis chosen at random from two possibilities, $\mathbb{X}$ and $\mathbb{Z}$, resulting in bit strings {\bf X} (for Alice) and {\bf X$'$} (for Bob). Next, they perform statistical tests on a few sample bits taken from {\bf X} and {\bf X$'$} in order to estimate the correlation. If this correlation is sufficiently large, they apply the above-mentioned post-processing procedure to turn their raw keys into a fully secret key of an appropriate length, $\ell$. Otherwise, if the estimated correlation is too small, they abort the protocol.

To prove that this protocol produces a secret key, it suffices to verify that the entropy difference in~\eqref{eqn:pp} is positive under the condition that the raw keys passed the correlation test. The second term of~\eqref{eqn:pp}, $\chmaxeps{{\bf X}}{{\bf X$'$}}{}$, directly depends on the correlation strength between the raw keys. For example, if {\bf X} and {\bf X$'$} consist of $n$ bits, of which at most a fraction $\delta$ disagree (according to the statistical test performed during the protocol), we have 
\begin{align}
  \label{eqn:maxbound}
  \chmaxeps{{\bf X}}{{\bf X$'$}}{} \lessapprox n h(\delta) \, ,
\end{align}
where $h(\cdot)$ denotes the binary entropy and $n$ is the number of bits in the raw key.

The first term in~\eqref{eqn:pp}, $\chmineps{{\bf X}}{E}{}$, depends on the correlations between {\bf X} and the adversary's information E, which is not accessible to Alice and Bob. The challenge is to bound these correlations from the data that is available, namely the correlations between {\bf X} and {\bf X$'$}. This is exactly where our uncertainty relation steps in.

Recall that, according to the protocol description, Alice and Bob measure each of their qubits with respect to one out of two different bases. One may now think of a hypothetical run of the protocol where Alice and Bob use the opposite basis choice for the measurement of each of their qubits, resulting in outcomes {\bf Y} and {\bf Y$'$}, respectively. We may then apply our uncertainty relation, which gives
\begin{align*}
  \chmineps{{\bf X}}{E}{} \geq q n - \chmaxeps{{\bf Y}}{{\bf Y$'$}}{} 
  = q n - \chmaxeps{\bf X}{\bf X$'$}{} \, ,
\end{align*}
where $q$ is evaluated for Alice's apparatus~\footnote{The parameter
  $q = -\log c$ is determined by the maximum overlap between the two
  bases used by Alice, e.g.~we have $q = 1$ for BB84. If the bases are
  not complementary to each other, then $q$ is reduced accordingly}. The last equality follows because the choice of basis was random for each qubit, and hence the correlation between {\bf Y} and {\bf Y$'$} is identical to the one between {\bf X} and {\bf X$'$}.  Inserting this into~\eqref{eqn:pp} and using~\eqref{eqn:maxbound}, we conclude that the protocol generates a secure key of length
\begin{align}
	\label{eqn:keylength}
  \ell \approx n \big( q - 2 h(\delta) \big) \, .
\end{align}

We emphasize that, in contrast to security proofs based on previous versions of the uncertainty relation, e.g.~\cite{koashi06} and~\cite{berta10}, this security proof does not rely on additional arguments such as the post-selection technique~\cite{christandl09}, the de Finetti theorem~\cite{renner07} and the quantum asymptotic equipartition property~\cite{renner05,tomamichel08}. Employing these tools introduces additional terms in~\eqref{eqn:keylength} that reduce the extractable key length significantly for experimentally feasible values of $n$. Our proof technique will therefore lead to tighter finite-key bounds~\cite{scarani08,sheridan10}. 

Finally, we note that our approach is different from recent
device-independent security proofs for entanglement-based
protocols~\cite{ekert91}, which are based on a violation of Bell's
theorem~\cite{haenggithesis10,masanes10}. In these proofs Assumption~3 applies to both parties and cannot be dropped|instead, it may be replaced by the
assumption that the measurement devices are memoryless.

\emph{Acknowledgments.}|\,We thank M.~Berta, M.~Christandl, R.~Colbeck and J.~Renes for many stimulating discussions. S.~Fehr pointed out an inadequacy in a previous version of the manuscript.
We acknowledge support from the Swiss National Science Foundation (grant No.\ 200021-119868).


\begin{thebibliography}{37}%
\makeatletter
\providecommand \@ifxundefined [1]{%
 \@ifx{#1\undefined}
}%
\providecommand \@ifnum [1]{%
 \ifnum #1\expandafter \@firstoftwo
 \else \expandafter \@secondoftwo
 \fi
}%
\providecommand \@ifx [1]{%
 \ifx #1\expandafter \@firstoftwo
 \else \expandafter \@secondoftwo
 \fi
}%
\providecommand \natexlab [1]{#1}%
\providecommand \enquote  [1]{``#1''}%
\providecommand \bibnamefont  [1]{#1}%
\providecommand \bibfnamefont [1]{#1}%
\providecommand \citenamefont [1]{#1}%
\providecommand \href@noop [0]{\@secondoftwo}%
\providecommand \href [0]{\begingroup \@sanitize@url \@href}%
\providecommand \@href[1]{\@@startlink{#1}\@@href}%
\providecommand \@@href[1]{\endgroup#1\@@endlink}%
\providecommand \@sanitize@url [0]{\catcode `\\12\catcode `\$12\catcode
  `\&12\catcode `\#12\catcode `\^12\catcode `\_12\catcode `\%12\relax}%
\providecommand \@@startlink[1]{}%
\providecommand \@@endlink[0]{}%
\providecommand \url  [0]{\begingroup\@sanitize@url \@url }%
\providecommand \@url [1]{\endgroup\@href {#1}{\urlprefix }}%
\providecommand \urlprefix  [0]{URL }%
\providecommand \arxivid [0]{\href }%
\@ifxundefined \urlstyle {%
  \providecommand \doi  [0]{\begingroup \@sanitize@url \@doi}%
  \providecommand \@doi [1]{\endgroup \@@startlink {\doibase
  #1}doi:\discretionary {}{}{}#1\@@endlink }%
}{%
  \providecommand \doi  [0]{doi:\discretionary{}{}{}\begingroup
  \urlstyle{rm}\Url }%
}%
\providecommand \doibase [0]{http://dx.doi.org/}%
\providecommand \Doi [0]{\begingroup \@sanitize@url \@Doi }%
\providecommand \@Doi  [1]{\endgroup\@@startlink{\doibase#1}\@@Doi}%
\providecommand \@@Doi [1]{#1\@@endlink}%
\providecommand \selectlanguage [0]{\@gobble}%
\providecommand \bibinfo  [0]{\@secondoftwo}%
\providecommand \bibfield  [0]{\@secondoftwo}%
\providecommand \translation [1]{[#1]}%
\providecommand \BibitemOpen [0]{}%
\providecommand \bibitemStop [0]{}%
\providecommand \bibitemNoStop [0]{.\EOS\space}%
\providecommand \EOS [0]{\spacefactor3000\relax}%
\providecommand \BibitemShut  [1]{\csname bibitem#1\endcsname}%
\bibitem [{\citenamefont {Heisenberg}(1927)}]{heisenberg27}%
  \BibitemOpen
  \bibfield  {author} {\bibinfo {author} {\bibfnamefont {W.}~\bibnamefont
  {Heisenberg}},\ }\href@noop {} {\bibfield  {journal} {\bibinfo  {journal} {Z.
  Phys.},\ }\textbf {\bibinfo {volume} {43}},\ \bibinfo {pages} {172} (\bibinfo
  {year} {1927})}\BibitemShut {NoStop}%
\bibitem [{\citenamefont {Maassen}\ and\ \citenamefont
  {Uffink}(1988)}]{maassen88}%
  \BibitemOpen
  \bibfield  {author} {\bibinfo {author} {\bibfnamefont {H.}~\bibnamefont
  {Maassen}}\ and\ \bibinfo {author} {\bibfnamefont {J.~B.~M.}~\bibnamefont
  {Uffink}},\ }\href@noop {} {\bibfield  {journal} {\bibinfo  {journal} {Phys.
  Rev. Lett.},\ }\textbf {\bibinfo {volume} {60}},\ \bibinfo {pages} {1103}
  (\bibinfo {year} {1988})}\BibitemShut {NoStop}%
\bibitem [{\citenamefont {Krishna}\ and\ \citenamefont
  {Parthasarathy}(2001)}]{krishna01}%
  \BibitemOpen
  \bibfield  {author} {\bibinfo {author} {\bibfnamefont {M.}~\bibnamefont
  {Krishna}}\ and\ \bibinfo {author} {\bibfnamefont {K.~R.}\ \bibnamefont
  {Parthasarathy}},\ }\href@noop {} {\bibfield  {journal} {\bibinfo  {journal}
  {Indian J. Stat.},\ }\textbf {\bibinfo {volume} {64}},\ \bibinfo {pages}
  {842} (\bibinfo {year} {2002})}\BibitemShut {NoStop}%
\bibitem [{\citenamefont {Renes}\ and\ \citenamefont
  {Boileau}(2009)}]{renes09}%
  \BibitemOpen
  \bibfield  {author} {\bibinfo {author} {\bibfnamefont {J.~M.}\ \bibnamefont
  {Renes}}\ and\ \bibinfo {author} {\bibfnamefont {J.-C.}\ \bibnamefont
  {Boileau}},\ }\href@noop {} {\bibfield  {journal} {\bibinfo  {journal} {Phys.
  Rev. Lett.},\ }\textbf {\bibinfo {volume} {103}},\ \bibinfo {pages} {020402}
  (\bibinfo {year} {2009})}\BibitemShut {NoStop}%
\bibitem [{Note1()}]{Note1}%
  \BibitemOpen
  \bibinfo {note} {The norm $|\protect \tmspace -\thinmuskip {.1667em}| \cdot
  |\protect \tmspace -\thinmuskip {.1667em}|_{\infty }$ evaluates the largest
  singular value. If the measurements are projective and rank 1, namely if
  $M_{x} = | x \protect \rangle \protect \tmspace -\thinmuskip {.1667em}
  \protect \langle x |$ and $N_z = | z \protect \rangle \protect \tmspace
  -\thinmuskip {.1667em} \protect \langle z |$, then~\protect \textup {\hbox
  {\mathsurround \z@ \protect \normalfont (\ignorespaces \ref
  {eqn:overlap}\unskip \@@italiccorr )}} reduces to the maximum overlap, $c =
  \protect \qopname \relax m{max}_{x, z} | \protect \langle x | z \protect
  \rangle |^2$}\BibitemShut {NoStop}%
\bibitem [{\citenamefont {Berta}\ \emph {et~al.}(2010)\citenamefont {Berta},
  \citenamefont {Christandl}, \citenamefont {Colbeck}, \citenamefont {Renes},\
  and\ \citenamefont {Renner}}]{berta10}%
  \BibitemOpen
  \bibfield  {author} {\bibinfo {author} {\bibfnamefont {M.}~\bibnamefont
  {Berta}}, \bibinfo {author} {\bibfnamefont {M.}~\bibnamefont {Christandl}},
  \bibinfo {author} {\bibfnamefont {R.}~\bibnamefont {Colbeck}}, \bibinfo
  {author} {\bibfnamefont {J.~M.}\ \bibnamefont {Renes}}, \ and\ \bibinfo
  {author} {\bibfnamefont {R.}~\bibnamefont {Renner}},\ }\href@noop {}
  {\bibfield  {journal} {\bibinfo  {journal} {Nat. Phys.},\ }\textbf {\bibinfo
  {volume} {6}},\ \bibinfo {pages} {659} (\bibinfo {year} {2010})}\BibitemShut
  {NoStop}%
\bibitem [{\citenamefont {Coles}\ \emph {et~al.}(2010)\citenamefont {Coles},
  \citenamefont {Yu}, \citenamefont {Gheorghiu},\ and\ \citenamefont
  {Griffiths}}]{coles10}%
  \BibitemOpen
  \bibfield  {author} {\bibinfo {author} {\bibfnamefont {P.~J.}\ \bibnamefont
  {Coles}}, \bibinfo {author} {\bibfnamefont {L.}~\bibnamefont {Yu}}, \bibinfo
  {author} {\bibfnamefont {V.}~\bibnamefont {Gheorghiu}}, \ and\ \bibinfo
  {author} {\bibfnamefont {R.~B.}\ \bibnamefont {Griffiths}}} (\bibinfo {year}
  {2010}),\ \arxivid {arXiv:\,http://arxiv.org/abs/1006.4859}
  {arXiv:\,1006.4859} \BibitemShut {NoStop}%
\bibitem [{Note2()}]{Note2}%
  \BibitemOpen
  \bibinfo {note} {Note that classical (unlike quantum) information can be
  ``copied'' and therefore be stored in two disjoint subsystems, B and
  C.}\BibitemShut {Stop}%
\bibitem [{Note3()}]{Note3}%
  \BibitemOpen
  \bibinfo {note} {We omit identity operators whenever their presence is
  implied by context, e.g.\ $M_x {\rho }_{\protect \textnormal {\relax \protect
  \fontsize {5}{6}\protect \selectfont ABC}}$ should be understood as $(M_x
  \otimes \protect \ensuremath {\protect \mathds {1}_{\protect \textnormal
  {\relax \protect \fontsize {5}{6}\protect \selectfont BC}}}) {\rho
  }_{\protect \textnormal {\relax \protect \fontsize {5}{6}\protect \selectfont
  ABC}}$.}\BibitemShut {Stop}%
\bibitem [{\citenamefont {Renner}\ and\ \citenamefont
  {Wolf}(2005)}]{rennerwolf05}%
  \BibitemOpen
  \bibfield  {author} {\bibinfo {author} {\bibfnamefont {R.}~\bibnamefont
  {Renner}}\ and\ \bibinfo {author} {\bibfnamefont {S.}~\bibnamefont {Wolf}},\
  }in\ \href@noop {} {\emph {\bibinfo {booktitle} {Advances in Cryptography ---
  ASIACRYPT}}},\ \bibinfo {series} {LNCS (Springer)}, Vol.\ \bibinfo {volume}
  {3788}\ (\bibinfo {year} {2005})\ pp.\ \bibinfo {pages}
  {199--216}\BibitemShut {NoStop}%
\bibitem [{\citenamefont {Renner}(2005)}]{renner05}%
  \BibitemOpen
  \bibfield  {author} {\bibinfo {author} {\bibfnamefont {R.}~\bibnamefont
  {Renner}},\ }\emph {\bibinfo {title} {{Security of Quantum Key
  Distribution}}},\ \href@noop {} {Ph.D. thesis},\ \bibinfo  {school} {ETH
  Zurich} (\bibinfo {year} {2005}),\ \arxivid
  {arXiv:\,http://arxiv.org/abs/quant-ph/0512258} {arXiv:\,quant-ph/0512258}
  \BibitemShut {NoStop}%
\bibitem [{Note4()}]{Note4}%
  \BibitemOpen
  \bibinfo {note} {Most results involving the von Neumann entropy are only
  valid for processes that produce a sequence of identical and independently
  distributed (i.i.d.) random values (see~\cite {renner05} for a
  discussion).}\BibitemShut {Stop}%
\bibitem [{\citenamefont {Wehner}\ and\ \citenamefont
  {Winter}(2009)}]{wehner09}%
  \BibitemOpen
  \bibfield  {author} {\bibinfo {author} {\bibfnamefont {S.}~\bibnamefont
  {Wehner}}\ and\ \bibinfo {author} {\bibfnamefont {A.}~\bibnamefont {Winter}}}
  (\bibinfo {year} {2009}),\ \arxivid {arXiv:\,http://arxiv.org/abs/0907.3704}
  {arXiv:\,0907.3704} \BibitemShut {NoStop}%
\bibitem [{\citenamefont {Renes}(2010)}]{renes10}%
  \BibitemOpen
  \bibfield  {author} {\bibinfo {author} {\bibfnamefont {J.~M.}\ \bibnamefont
  {Renes}}} (\bibinfo {year} {2010}),\ \arxivid
  {arXiv:\,http://arxiv.org/abs/1003.0703} {arXiv:\,1003.0703} \BibitemShut
  {NoStop}%
\bibitem [{\citenamefont {Tomamichel}\ \emph {et~al.}(2009)\citenamefont
  {Tomamichel}, \citenamefont {Colbeck},\ and\ \citenamefont
  {Renner}}]{tomamichel08}%
  \BibitemOpen
  \bibfield  {author} {\bibinfo {author} {\bibfnamefont {M.}~\bibnamefont
  {Tomamichel}}, \bibinfo {author} {\bibfnamefont {R.}~\bibnamefont {Colbeck}},
  \ and\ \bibinfo {author} {\bibfnamefont {R.}~\bibnamefont {Renner}},\
  }\href@noop {} {\bibfield  {journal} {\bibinfo  {journal} {IEEE Trans. on
  Inf. Theory},\ }\textbf {\bibinfo {volume} {55}},\ \bibinfo {pages} {5840}
  (\bibinfo {year} {2009})}\BibitemShut {NoStop}%
\bibitem [{\citenamefont {R\'{e}nyi}(1961)}]{renyi61}%
  \BibitemOpen
  \bibfield  {author} {\bibinfo {author} {\bibfnamefont {A.}~\bibnamefont
  {R\'{e}nyi}},\ }in\ \href@noop {} {\emph {\bibinfo {booktitle} {Proc. Symp.
  on Math., Stat. and Probability}}}\ (\bibinfo {address} {Berkeley},\ \bibinfo
  {year} {1961})\ pp.\ \bibinfo {pages} {547--561}\BibitemShut {NoStop}%
\bibitem [{\citenamefont {Damgaard}\ \emph {et~al.}(2007)\citenamefont
  {Damgaard}, \citenamefont {Fehr}, \citenamefont {Renner}, \citenamefont
  {Salvail},\ and\ \citenamefont {Schaffner}}]{damgaard07}%
  \BibitemOpen
  \bibfield  {author} {\bibinfo {author} {\bibfnamefont {I.~B.}\ \bibnamefont
  {Damgaard}}, \bibinfo {author} {\bibfnamefont {S.}~\bibnamefont {Fehr}},
  \bibinfo {author} {\bibfnamefont {R.}~\bibnamefont {Renner}}, \bibinfo
  {author} {\bibfnamefont {L.}~\bibnamefont {Salvail}}, \ and\ \bibinfo
  {author} {\bibfnamefont {C.}~\bibnamefont {Schaffner}},\ }in\ \href@noop {}
  {\emph {\bibinfo {booktitle} {Advances in Cryptography --- CRYPTO}}},\
  \bibinfo {series} {LNCS (Springer)}, Vol.\ \bibinfo {volume} {4622}\
  (\bibinfo {year} {2007})\ pp.\ \bibinfo {pages} {360--378}\BibitemShut
  {NoStop}%
\bibitem [{\citenamefont {Bennett}\ and\ \citenamefont
  {Brassard}(1984)}]{bb84}%
  \BibitemOpen
  \bibfield  {author} {\bibinfo {author} {\bibfnamefont {C.~H.}\ \bibnamefont
  {Bennett}}\ and\ \bibinfo {author} {\bibfnamefont {G.}~\bibnamefont
  {Brassard}},\ }in\ \href@noop {} {\emph {\bibinfo {booktitle} {Proc. IEEE
  Int. Conf. on Comp., Sys. and Signal Process.}}}\ (\bibinfo {address}
  {Bangalore},\ \bibinfo {year} {1984})\ pp.\ \bibinfo {pages}
  {175--179}\BibitemShut {NoStop}%
\bibitem [{\citenamefont {Ekert}(1991)}]{ekert91}%
  \BibitemOpen
  \bibfield  {author} {\bibinfo {author} {\bibfnamefont {A.~K.}\ \bibnamefont
  {Ekert}},\ }\href@noop {} {\bibfield  {journal} {\bibinfo  {journal} {Phys.
  Rev. Lett.},\ }\textbf {\bibinfo {volume} {67}},\ \bibinfo {pages} {661}
  (\bibinfo {year} {1991})}\BibitemShut {NoStop}%
\bibitem [{\citenamefont {Tomamichel}\ \emph {et~al.}(2010)\citenamefont
  {Tomamichel}, \citenamefont {Colbeck},\ and\ \citenamefont
  {Renner}}]{tomamichel09}%
  \BibitemOpen
  \bibfield  {author} {\bibinfo {author} {\bibfnamefont {M.}~\bibnamefont
  {Tomamichel}}, \bibinfo {author} {\bibfnamefont {R.}~\bibnamefont {Colbeck}},
  \ and\ \bibinfo {author} {\bibfnamefont {R.}~\bibnamefont {Renner}},\
  }\href@noop {} {\bibfield  {journal} {\bibinfo  {journal} {IEEE Trans. on
  Inf. Theory},\ }\textbf {\bibinfo {volume} {56}},\ \bibinfo {pages} {4674}
  (\bibinfo {year} {2010})}\BibitemShut {NoStop}%
\bibitem [{\citenamefont {K\"{o}nig}\ \emph {et~al.}(2009)\citenamefont
  {K\"{o}nig}, \citenamefont {Renner},\ and\ \citenamefont
  {Schaffner}}]{koenig08}%
  \BibitemOpen
  \bibfield  {author} {\bibinfo {author} {\bibfnamefont {R.}~\bibnamefont
  {K\"{o}nig}}, \bibinfo {author} {\bibfnamefont {R.}~\bibnamefont {Renner}}, \
  and\ \bibinfo {author} {\bibfnamefont {C.}~\bibnamefont {Schaffner}},\
  }\href@noop {} {\bibfield  {journal} {\bibinfo  {journal} {IEEE Trans. on
  Inf. Theory},\ }\textbf {\bibinfo {volume} {55}},\ \bibinfo {pages} {4337}
  (\bibinfo {year} {2009})}\BibitemShut {NoStop}%
\bibitem [{\citenamefont {Lydersen}\ \emph {et~al.}(2010)\citenamefont
  {Lydersen}, \citenamefont {Wiechers}, \citenamefont {Wittmann}, \citenamefont
  {Elser}, \citenamefont {Skaar},\ and\ \citenamefont {Makarov}}]{lydersen10}%
  \BibitemOpen
  \bibfield  {author} {\bibinfo {author} {\bibfnamefont {L.}~\bibnamefont
  {Lydersen}}, \bibinfo {author} {\bibfnamefont {C.}~\bibnamefont {Wiechers}},
  \bibinfo {author} {\bibfnamefont {C.}~\bibnamefont {Wittmann}}, \bibinfo
  {author} {\bibfnamefont {D.}~\bibnamefont {Elser}}, \bibinfo {author}
  {\bibfnamefont {J.}~\bibnamefont {Skaar}}, \ and\ \bibinfo {author}
  {\bibfnamefont {V.}~\bibnamefont {Makarov}},\ }\href
  {http://dx.doi.org/10.1038/nphoton.2010.214} {\bibfield  {journal} {\bibinfo
  {journal} {Nat. Photonics},\ }\textbf {\bibinfo {volume} {4}},\ \bibinfo
  {pages} {686} (\bibinfo {year} {2010})}\BibitemShut {NoStop}%
\bibitem [{\citenamefont {Xu}\ \emph {et~al.}(2010)\citenamefont {Xu},
  \citenamefont {Qi},\ and\ \citenamefont {Lo}}]{lo10}%
  \BibitemOpen
  \bibfield  {author} {\bibinfo {author} {\bibfnamefont {F.}~\bibnamefont
  {Xu}}, \bibinfo {author} {\bibfnamefont {B.}~\bibnamefont {Qi}}, \ and\
  \bibinfo {author} {\bibfnamefont {H.-K.}\ \bibnamefont {Lo}},\ }\href@noop {}
  {\bibfield  {journal} {\bibinfo  {journal} {New Journal of Physics},\
  }\textbf {\bibinfo {volume} {12}},\ \bibinfo {pages} {113026} (\bibinfo
  {year} {2010})}\BibitemShut {NoStop}%
\bibitem [{\citenamefont {Mayers}(1996)}]{mayers96}%
  \BibitemOpen
  \bibfield  {author} {\bibinfo {author} {\bibfnamefont {D.}~\bibnamefont
  {Mayers}},\ }in\ \href@noop {} {\emph {\bibinfo {booktitle} {Advances in
  Cryptography --- CRYPTO}}},\ \bibinfo {series} {LNCS (Springer)}, Vol.\
  \bibinfo {volume} {1109}\ (\bibinfo {year} {1996})\ pp.\ \bibinfo {pages}
  {343--357}\BibitemShut {NoStop}%
\bibitem [{\citenamefont {Bennett}\ \emph {et~al.}(1995)\citenamefont
  {Bennett}, \citenamefont {Brassard}, \citenamefont {Crepeau},\ and\
  \citenamefont {Maurer}}]{bennett95}%
  \BibitemOpen
  \bibfield  {author} {\bibinfo {author} {\bibfnamefont {C.~H.}\ \bibnamefont
  {Bennett}}, \bibinfo {author} {\bibfnamefont {G.}~\bibnamefont {Brassard}},
  \bibinfo {author} {\bibfnamefont {C.}~\bibnamefont {Crepeau}}, \ and\
  \bibinfo {author} {\bibfnamefont {U.~M.}\ \bibnamefont {Maurer}},\
  }\href@noop {} {\bibfield  {journal} {\bibinfo  {journal} {IEEE Trans. on
  Inf. Theory},\ }\textbf {\bibinfo {volume} {41}},\ \bibinfo {pages} {1915}
  (\bibinfo {year} {1995})}\BibitemShut {NoStop}%
\bibitem [{\citenamefont {Renner}\ and\ \citenamefont
  {K\"{o}nig}(2005)}]{rennerkoenig05}%
  \BibitemOpen
  \bibfield  {author} {\bibinfo {author} {\bibfnamefont {R.}~\bibnamefont
  {Renner}}\ and\ \bibinfo {author} {\bibfnamefont {R.}~\bibnamefont
  {K\"{o}nig}},\ }in\ \href@noop {} {\emph {\bibinfo {booktitle} {Proc.
  TCC}}},\ \bibinfo {series} {LNCS (Springer)}, Vol.\ \bibinfo {volume} {3378}\
  (\bibinfo {address} {Cambridge, USA},\ \bibinfo {year} {2005})\ pp.\ \bibinfo
  {pages} {407--425}\BibitemShut {NoStop}%
\bibitem [{\citenamefont {Renes}\ and\ \citenamefont
  {Renner}(2010)}]{renesrenner10}%
  \BibitemOpen
  \bibfield  {author} {\bibinfo {author} {\bibfnamefont {J.~M.}\ \bibnamefont
  {Renes}}\ and\ \bibinfo {author} {\bibfnamefont {R.}~\bibnamefont {Renner}}}
  (\bibinfo {year} {2010}),\ \arxivid {arXiv:\,http://arxiv.org/abs/1008.0452}
  {arXiv:\,1008.0452} \BibitemShut {NoStop}%
\bibitem [{\citenamefont {Devetak}\ and\ \citenamefont
  {Winter}(2005)}]{devetak05}%
  \BibitemOpen
  \bibfield  {author} {\bibinfo {author} {\bibfnamefont {I.}~\bibnamefont
  {Devetak}}\ and\ \bibinfo {author} {\bibfnamefont {A.}~\bibnamefont
  {Winter}},\ }\href@noop {} {\bibfield  {journal} {\bibinfo  {journal} {Proc.
  R. Soc. Lond. Ser. A Math. Phys. Eng. Sci.},\ }\textbf {\bibinfo {volume}
  {461}},\ \bibinfo {pages} {207} (\bibinfo {year} {2005})}\BibitemShut
  {NoStop}%
\bibitem [{\citenamefont {Bennett}\ \emph {et~al.}(1992)\citenamefont
  {Bennett}, \citenamefont {Brassard},\ and\ \citenamefont
  {Mermin}}]{bennett92}%
  \BibitemOpen
  \bibfield  {author} {\bibinfo {author} {\bibfnamefont {C.~H.}\ \bibnamefont
  {Bennett}}, \bibinfo {author} {\bibfnamefont {G.}~\bibnamefont {Brassard}}, \
  and\ \bibinfo {author} {\bibfnamefont {N.~D.}~\bibnamefont {Mermin}},\
  }\href@noop {} {\bibfield  {journal} {\bibinfo  {journal} {Phys. Rev.
  Lett.},\ }\textbf {\bibinfo {volume} {68}},\ \bibinfo {pages} {557} (\bibinfo
  {year} {1992})}\BibitemShut {NoStop}%
\bibitem [{Note5()}]{Note5}%
  \BibitemOpen
  \bibinfo {note} {The parameter $q = -\protect \qopname \relax o{log}c$ is
  determined by the maximum overlap between the two bases used by Alice,
  e.g.~we have $q = 1$ for BB84. If the bases are not complementary to each
  other, then $q$ is reduced accordingly}\BibitemShut {NoStop}%
\bibitem [{\citenamefont {Koashi}(2006)}]{koashi06}%
  \BibitemOpen
  \bibfield  {author} {\bibinfo {author} {\bibfnamefont {M.}~\bibnamefont
  {Koashi}},\ }\href@noop {} {\bibfield  {journal} {\bibinfo  {journal} {J.
  Phys. Conf. Ser.},\ }\textbf {\bibinfo {volume} {36}},\ \bibinfo {pages} {98}
  (\bibinfo {year} {2006})}\BibitemShut {NoStop}%
\bibitem [{\citenamefont {Christandl}\ \emph {et~al.}(2009)\citenamefont
  {Christandl}, \citenamefont {K\"{o}nig},\ and\ \citenamefont
  {Renner}}]{christandl09}%
  \BibitemOpen
  \bibfield  {author} {\bibinfo {author} {\bibfnamefont {M.}~\bibnamefont
  {Christandl}}, \bibinfo {author} {\bibfnamefont {R.}~\bibnamefont
  {K\"{o}nig}}, \ and\ \bibinfo {author} {\bibfnamefont {R.}~\bibnamefont
  {Renner}},\ }\href@noop {} {\bibfield  {journal} {\bibinfo  {journal} {Phys.
  Rev. Lett.},\ }\textbf {\bibinfo {volume} {102}},\ \bibinfo {pages} {020504}
  (\bibinfo {year} {2009})}\BibitemShut {NoStop}%
\bibitem [{\citenamefont {Renner}(2007)}]{renner07}%
  \BibitemOpen
  \bibfield  {author} {\bibinfo {author} {\bibfnamefont {R.}~\bibnamefont
  {Renner}},\ }\href@noop {} {\bibfield  {journal} {\bibinfo  {journal} {Nat.
  Phys.},\ }\textbf {\bibinfo {volume} {3}},\ \bibinfo {pages} {645} (\bibinfo
  {year} {2007})}\BibitemShut {NoStop}%
\bibitem [{\citenamefont {Scarani}\ and\ \citenamefont
  {Renner}(2008)}]{scarani08}%
  \BibitemOpen
  \bibfield  {author} {\bibinfo {author} {\bibfnamefont {V.}~\bibnamefont
  {Scarani}}\ and\ \bibinfo {author} {\bibfnamefont {R.}~\bibnamefont
  {Renner}},\ }\href@noop {} {\bibfield  {journal} {\bibinfo  {journal} {Phys.
  Rev. Lett.},\ }\textbf {\bibinfo {volume} {100}},\ \bibinfo {pages} {200501}
  (\bibinfo {year} {2008})}\BibitemShut {NoStop}%
\bibitem [{\citenamefont {Sheridan}\ \emph {et~al.}(2010)\citenamefont
  {Sheridan}, \citenamefont {Le},\ and\ \citenamefont {Scarani}}]{sheridan10}%
  \BibitemOpen
  \bibfield  {author} {\bibinfo {author} {\bibfnamefont {L.}~\bibnamefont
  {Sheridan}}, \bibinfo {author} {\bibfnamefont {T.~P.}\ \bibnamefont {Le}}, \
  and\ \bibinfo {author} {\bibfnamefont {V.}~\bibnamefont {Scarani}},\
  }\href@noop {} {\bibfield  {journal} {\bibinfo  {journal} {New J. Phys.},\
  }\textbf {\bibinfo {volume} {12}},\ \bibinfo {pages} {123019} (\bibinfo
  {year} {2010})}\BibitemShut {NoStop}%
\bibitem [{\citenamefont {H\"{a}nggi}(2010)}]{haenggithesis10}%
  \BibitemOpen
  \bibfield  {author} {\bibinfo {author} {\bibfnamefont {E.}~\bibnamefont
  {H\"{a}nggi}},\ }\emph {\bibinfo {title} {{Device-Independent Quantum Key
  Distribution}}},\ \href@noop {} {Ph.D. thesis},\ \bibinfo  {school} {ETH
  Zurich} (\bibinfo {year} {2010})\BibitemShut {NoStop}%
\bibitem [{\citenamefont {Masanes}\ \emph {et~al.}(2010)\citenamefont
  {Masanes}, \citenamefont {Pironio},\ and\ \citenamefont
  {Ac\'{\i}n}}]{masanes10}%
  \BibitemOpen
  \bibfield  {author} {\bibinfo {author} {\bibfnamefont {L.}~\bibnamefont
  {Masanes}}, \bibinfo {author} {\bibfnamefont {S.}~\bibnamefont {Pironio}}, \
  and\ \bibinfo {author} {\bibfnamefont {A.}~\bibnamefont {Ac\'{\i}n}},\ }\href
  {http://arxiv.org/abs/1009.1567} { (\bibinfo {year} {2010})},\ \arxivid
  {arXiv:\,http://arxiv.org/abs/1009.1567} {arXiv:\,1009.1567} \BibitemShut
  {NoStop}%
\end{thebibliography}

%

\end{document}